\documentstyle[aps,prl,psfig,multicol]{revtex}
\begin{document}
\draft
\title{Sub-Poissonian shot noise in non-degenerate diffusive conductors}
\author{C. W. J. Beenakker}
\address{Instituut-Lorentz, Leiden University,
P.O. Box 9506, 2300 RA Leiden, The Netherlands}
\date{submitted 9 October 1998, revised 23 November 1998}
\maketitle
\begin{abstract}
A theory is presented for the universal reduction of shot noise by Coulomb
repulsion, which was observed in computer simulations of a disordered
non-degenerate electron gas by Gonz\'{a}lez {\em et al.\/} [Phys.\ Rev.\ Lett.\
{\bf 80}, 2901 (1998)]. The universality of the reduction below the
uncorrelated value is explained as a feature of the high-voltage regime of
space-charge limited conduction. The reduction factor depends on the
dimensionality $d$ of the density of states, being close but not quite equal to
$1/d$ in two and three dimensions.
\end{abstract}
\pacs{PACS numbers: 72.70.+m, 72.20.Ht, 73.50.Fq, 73.50.Td}
\begin{multicols}{2}

The motivation for this work comes from a remarkable recent Letter \cite{Gon98}
by Gonz\'{a}lez, Gonz\'{a}lez, Mateos, Pardo, Reggiani, Bulashenko, and
Rub\'{\i} on the ``universality of the 1/3 shot-noise suppression factor in
non-degenerate diffusive conductors''. Shot noise is the time-dependent
fluctuation in the electrical current caused by the discreteness of the charge.
In the last few years there has been a breakthrough in the use of shot-noise
measurements to study correlation effects in diffusive conductors
\cite{reviews}. In the absence of correlations between the electrons, the
current fluctuations $\delta I(t)$ around the mean current $\bar{I}$ are
described by a Poisson process, with a spectral density $P$ at low frequencies
equal to $P_{\rm Poisson}=2e\bar{I}$. Correlations reduce the noise below the
Poisson value.

The Pauli exclusion principle is one source of correlations, Coulomb repulsion
is another. In a {\em degenerate\/} electron gas with elastic impurity
scattering the reduction is a factor one-third \cite{Bee92,Nag92}. This
reduction is due to the Pauli principle. The remarkable finding of Gonz\'{a}lez
{\em et al.\/} was that Coulomb repulsion in a three-dimensional {\em
non-degenerate\/} electron gas also gives $P/P_{\rm Poisson}=\frac{1}{3}$. They
argued for a universal physical principle behind the one-third reduction of the
shot noise from elastic scattering, regardless of whether the origin of the
electron correlations is quantum mechanical (Pauli principle) or classical
(Coulomb repulsion).

The significance of Ref.\ \cite{Gon98} has been assessed critically by Landauer
\cite{Lan98}. While in a degenerate electron gas the one-third reduction is
independent of the number $d$ of spatial dimensions, Ref.\ \cite{Gon98} finds
$P/P_{\rm Poisson}=\frac{1}{2}$ for $d=2$ --- spoiling the supposed
universality of the reduction factor in non-degenerate conductors. Still, it
remains a remarkable finding that the ratio $P/P_{\rm Poisson}$ has no
dependence on microscopic parameters (such as mean free path $l$ or dielectric
constant $\kappa$) or global parameters (such as temperature $T$, voltage $V$,
or sample length $L$), as long as one stays in the high-voltage, diffusive
regime ($eV\gg kT$ and $L\gg l$). The findings of Ref.\ \cite{Gon98} are based
on a computer simulation of the dynamics of an interacting electron gas. Here
we develop an analytical theory that explains these numerical results.

We use the same theoretical framework as Nagaev \cite{Nag92} used for the
degenerate electron gas, namely the Boltzmann-Langevin equation \cite{Kog96} in
the diffusion approximation (valid for $L\gg l$). A difference with Ref.\
\cite{Nag92} is that the kinetic energy $\varepsilon=\frac{1}{2}mv^{2}$ now
appears as an independent variable. (In the degenerate case one may assume that
all non-equilibrium electrons have velocity $v$ equal to the Fermi velocity.)
The charge density $\rho({\bf r},\varepsilon,t)=\bar\rho+\delta\rho$, current
density ${\bf j}({\bf r},\varepsilon,t)={\bf j}+\delta{\bf j}$, and electric
field ${\bf E}({\bf r},t)=\bar{\bf E}+\delta{\bf E}$ fluctuate in time around
their time-averaged values (indicated by an overline). In the low-frequency
regime of interest we may neglect the time-derivative in the continuity
equation,
\begin{equation}
\left(\frac{\partial}{\partial{\bf r}}+e{\bf
E}\frac{\partial}{\partial\varepsilon}\right)\cdot{\bf j}=0.\label{continuity}
\end{equation}

Current and charge density are related by the drift-diffusion equation
\begin{equation}
{\bf j}=-D\frac{\partial\rho}{\partial{\bf r}}-\sigma\frac{\partial
f}{\partial\varepsilon}{\bf E}+\delta{\bf J},\label{driftdiffusion}
\end{equation}
with $\sigma(\varepsilon)=e^{2}D(\varepsilon)\nu(\varepsilon)$ the
conductivity, $D(\varepsilon)=D_{0}\varepsilon$ the diffusion constant, and
$\nu(\varepsilon)=\nu_{0}\varepsilon^{d/2-1}$ the density of states. (The
coefficients $D_{0}$ and $\nu_{0}$ are $\varepsilon$-independent, assuming an
energy-independent effective mass and scattering rate.) The function $f({\bf
r},\varepsilon,t)=\rho/e\nu=\bar{f}+\delta f$ is the occupation number of a
quantum state. (In equilibrium, the mean $\bar{f}$ is the Fermi-Dirac
distribution function.) The ``Langevin current'' $\delta{\bf J}({\bf
r},\varepsilon,t)$ is a stochastic source of current fluctuations from elastic
scattering \cite{Kog96}. Its first two moments are $\overline{\delta{\bf J}}=0$
and
\begin{eqnarray}
\overline{\delta J_{i}({\bf r},\varepsilon,t)\delta J_{j}({\bf
r}',\varepsilon',t')}=2\sigma(\varepsilon)\bar{f}({\bf
r},\varepsilon)[1-\bar{f}({\bf r},\varepsilon)]\nonumber\\
\mbox{}\times\delta_{ij}\delta({\bf r}-{\bf
r}')\delta(\varepsilon-\varepsilon')\delta(t-t'). \label{variancedeltaJ}
\end{eqnarray}
The definition of a non-degenerate electron gas is $\bar{f}\ll 1$, so that we
may ignore the factor $1-\bar{f}$ in this correlator.

We need one more equation to close the problem, namely the Poisson equation
\begin{equation}
\kappa\frac{\partial}{\partial{\bf r}}\cdot{\bf E}=\int
d\varepsilon\,\left(\rho-\rho_{\rm eq}\right),\label{Poisson}
\end{equation}
with $\rho_{\rm eq}$ the compensating charge present in the semiconductor in
equilibrium (equal to the charge density of the carriers prior to the injection
from the contacts).

The geometry we are considering is a disordered semiconductor of uniform
cross-sectional area $A$ sandwiched between metal contacts at $x=0$ and $x=L$.
We denote by ${\bf r}_{\perp}$ the $d-1$ dimensional vector of transverse
coordinates. (Only $d=3$ is physically relevant, but we consider arbitrary $d$
for comparison with the computer simulations.) The mean values of $\rho$, ${\bf
j}$, and ${\bf E}$ are independent of ${\bf r}_{\perp}$, but the fluctuations
are not. We define the linear charge density $\rho(x,t)=\int d{\bf
r}_{\perp}\int d\varepsilon\,\rho({\bf r},\varepsilon,t)$, the electric field
profile $E(x,t)=A^{-1}\int d{\bf r}_{\perp}\,E_{x}({\bf r},t)$, and the
currents $I(t)=\int d{\bf r}_{\perp}\int d\varepsilon\,j_{x}({\bf
r},\varepsilon,t)$ and $\delta J(x,t)=\int d{\bf r}_{\perp}\int
d\varepsilon\,\delta J_{x}({\bf r},\varepsilon,t)$. The total current $I$ is
independent of $x$ at low frequencies because of the continuity equation, but
the Langevin current $\delta J$ is not so restricted. In view of Eq.\
(\ref{driftdiffusion}), the two currents are related by
\begin{equation}
I=-\int d{\bf r}_{\perp}\int d\varepsilon\,D\frac{\partial\rho}{\partial
x}+\mu\rho E+\delta J,\label{IJrelation}
\end{equation}
with $\mu=\frac{1}{2}deD_{0}$ the mobility. (To write the drift term in the
form $\mu\rho E$ we have made a partial integration over energy and linearized
with respect to the fluctuations.)

A non-fluctuating voltage $V$ is applied between the two metal contacts, with
the current source at $x=0$ and the current drain at $x=L$. (The charge $e$ of
the carriers is taken to be positive.) For high $V$ the charge injected into
the semiconductor by the current source will be much larger than the charge
$\rho_{\rm eq}$ present in equilibrium. We will neglect $\rho_{\rm eq}$
altogether. For sufficiently high voltages, when all the surface charge at
$x=0$ has been injected into the semiconductor, the system enters the regime of
space-charged limited conduction, characterized by the boundary condition
\begin{equation}
\bar{E}(x)=0\;\;{\rm at}\;\;x=0.\label{boundarycondition}
\end{equation}
The mean charge and field distributions in this regime were studied extensively
in the past \cite{Lam70}, but apparently the shot-noise problem was not. We
argue that the universality of the computer simulations \cite{Gon98} is a
consequence of the homogeneity of the boundary condition
(\ref{boundarycondition}). Indeed, if the boundary condition would have
contained an external electric field, then the effect of Coulomb repulsion on
the shot noise would have depended on the relative magnitude of the induced and
external fields and hence on the value of $\kappa$. No universal reduction
factor could have resulted. This scenario stands opposite to that in the
degenerate case. There the reduction of shot noise occurs at low voltages, in
the linear-response regime, when the induced electric field can be neglected
relative to the external field \cite{note1}.

The zero-frequency limit of the noise spectral density is given by
\begin{equation}
P=2\int_{-\infty}^{\infty}dt\,\overline{\delta I(0)\delta I(t)}.\label{Pdef}
\end{equation}
To compute $P$ we need to relate the correlator of the total current $\delta I$
to the correlator of the Langevin current $\delta J$. At non-zero temperatures,
$\delta I$ contains also a contribution from the thermal fluctuations of the
charge at the contacts (Johnson-Nyquist noise \cite{Kog96}). This source of
noise may be neglected relative to the shot noise for $eV\gg kT$, and we will
do so to simplify the problem. The most questionable simplification that we
will make is to neglect the diffusion term ($\propto\partial\rho/\partial x$)
relative to the drift term ($\propto E$) in the drift-diffusion equation
(\ref{IJrelation}). This approximation is customary in treatments of
space-charge limited conduction \cite{Lam70}, but is only rigorously justified
here in the formal limit $d\rightarrow\infty$ (when the ratio
$\mu/D_{0}\rightarrow\infty$).

We are now ready to proceed to a solution of the coupled kinetic and Poisson
equations. We consider first the mean values and then the fluctuations.
Combination of Eq.\ (\ref{Poisson}) (without the term $\rho_{\rm eq}$) and Eq.\
(\ref{IJrelation}) (without the diffusion term) gives for the mean electric
field
\begin{equation}
\bar{E}\frac{d\bar{E}}{dx}=\frac{\bar{I}}{\mu\kappa A}\Rightarrow\bar{E}(x)=
\left(\frac{2\bar{I}x}{\mu\kappa A}\right)^{1/2},\label{MottGurney}
\end{equation}
where we have used the boundary condition (\ref{boundarycondition}). The
$\surd{x}$ dependence of the electric field is the celebrated Mott-Gurney law
\cite{Mot40}. The corresponding charge density has an inverse square-root
singularity at $x=0$ \cite{note3}. The corresponding voltage
$V=\int_{0}^{L}\bar{E}\,dx\propto\surd{\bar{I}}$, so that the current increases
quadratically with the voltage. These are well-known results for space-charge
limited conduction \cite{Lam70}.

Linearization of Eqs.\ (\ref{Poisson}) and (\ref{IJrelation}) around the mean
values gives for the fluctuations
\begin{eqnarray}
&&\bar{E}\frac{\partial}{\partial x}\delta E+\frac{d\bar{E}}{dx}\delta
E=\frac{\delta I-\delta J}{\mu\kappa A}\nonumber\\
&&\mbox{}\Rightarrow\delta E(x,t)=x^{-1/2}\int_{0}^{x}dx'\,\frac{\delta
I(t)-\delta J(x',t)}{(2\bar{I}\mu\kappa A)^{1/2}}.\label{deltaEequation}
\end{eqnarray}
A non-fluctuating voltage requires $\int_{0}^{L}\delta E\,dx=0$, hence
\begin{equation}
\delta I(t)=3\int_{0}^{L}\frac{dx}{L}\,\Bigl(1-\sqrt{x/L}\Bigr)\delta
J(x,t).\label{deltaIdeltaJ}
\end{equation}
Combination of this relation between $\delta I$ and $\delta J$ with Eqs.\
(\ref{variancedeltaJ}) and (\ref{Pdef}) gives an expression for the shot-noise
power,
\begin{equation}
P=\frac{36A}{L}\int_{0}^{L}\frac{dx}{L}\Bigl(1-\sqrt{x/L}\Bigr)^{2}\int
d\varepsilon\,\sigma(\varepsilon)\bar{f}(x,\varepsilon).\label{Pbarf}
\end{equation}

To evaluate this expression we need to know the mean occupation number
$\bar{f}$ out of equilibrium. For this purpose it is convenient to change
variables from kinetic energy $\varepsilon$ to total energy
$u=\varepsilon+e\phi(x)$, with $\phi(x)$ the mean electrical potential. Since
$\bar{E}=-d\phi/dx$, the derivative $\partial/\partial
x+e\bar{E}\,\partial/\partial\varepsilon$ is equivalent to $\partial/\partial
x$ at constant $u$. The kinetic equations (\ref{continuity}),
(\ref{driftdiffusion}) in the new variables $x,u$ take the form
\begin{equation}
\frac{\partial\bar{j}}{\partial
x}=0,\;\;\bar{j}=-\frac{1}{e}\sigma[u-e\phi(x)]\frac{\partial\bar{f}}{\partial
x}.\label{xuequation}
\end{equation}
The solution is
\begin{equation}
\bar{f}(x,u)=\frac{\bar{f}(0,u)}{AR(u)}\int_{x}^{L}
\frac{dx'}{\sigma[u-e\phi(x')]}, \label{fbarxu}
\end{equation}
where we have imposed the absorbing boundary condition $\bar{f}(L,u)=0$ at the
current drain. (At high voltages the charge injected into the semiconductor by
the current drain can be neglected.) The factor
$R(u)=A^{-1}\int_{0}^{L}dx/\sigma[u-e\phi(x)]$ is the resistance of the
semiconductor. The mean current is related to $R$ by $e\bar{I}=\int
du\,\bar{f}(0,u)/R(u)$. The argument $u-e\phi(x)$ of $\sigma$ may be replaced
by $eV(x/L)^{3/2}$ in the high-$V$ limit. Then $\sigma\propto x^{3d/4}$ and
\begin{eqnarray}
\int du\,\sigma[u-e\phi(x)]\bar{f}(x,u)\rightarrow
\frac{e\bar{I}}{A}\int_{x}^{L}dx'\,\left(\frac{x}{x'}\right)^{3d/4}\nonumber\\
=\frac{4e\bar{I}L}{(3d-4)A}\left[\frac{x}{L}-
\left(\frac{x}{L}\right)^{3d/4}\right].\label{sigmafresult}
\end{eqnarray}

Substitution into Eq.\ (\ref{Pbarf}) yields our final result
\begin{eqnarray}
P&=&\frac{144e\bar{I}}{3d-4}\int_{0}^{1}dx\,
(1-\sqrt{x})^{2}(x-x^{3d/4})\nonumber\\
&=&\frac{24e\bar{I}}{5}\,\frac{3d^{2}+22d+64}{(d+2)(3d+4)(3d+8)}.\label{Pfinal}
\end{eqnarray}
The ratio $P/P_{\rm Poisson}$ equals 0.341 and 0.514 for $d=3$ and $d=2$,
respectively, within error bars of the fractions $\frac{1}{3}$ and
$\frac{1}{2}$ inferred by Gonz\'{a}lez {\em et al.\/} from their computer
simulations \cite{Gon98}. The proximity of these numbers to $d^{-1}$ appears to
be accidental. Indeed, for large $d$ we find that $P/P_{\rm
Poisson}\rightarrow\frac{4}{5}d^{-1}$. The large-$d$ limit is a rigorous
result, while the finite-$d$ values are not because we have neglected the
diffusion term in Eq.\ (\ref{IJrelation}).

In closing, we comment on the universality of the results and on their
experimental observability. Concerning the universality, the dimensionality
dependence has already been noted \cite{Gon98}. For a given $d$ there is no
dependence on material parameters, however, the shot noise does depend on the
model chosen for the energy-dependence of the elastic scattering rate. We have
followed the computer simulations \cite{Gon98} in assuming an
$\varepsilon$-independent scattering rate. In a model of short-range impurity
scattering one would have instead a rate proportional to the density of states.
This would change the energy dependence of the diffusion constant from
$D\propto\varepsilon$ to $D\propto\varepsilon^{2-d/2}$. The shot-noise power
remains unaffected for $d=2$, but for $d=3$ one obtains \cite{Sch98} $P/P_{\rm
Poisson}=0.407$ --- some 20\% above the value for an $\varepsilon$-independent
scattering rate. Concerning the experimental observability, the main obstacle
is the tendency of electron-phonon scattering to equilibrate the electron gas
at the lattice temperature. Then instead of shot noise one would measure
thermal noise (modified for a non-Ohmic conductor \cite{note2}), that is not
sensitive to correlation effects. Experiments in degenerate systems have
succeeded recently in observing shot noise by reducing the sample dimensions to
the mesoscopic scale \cite{experiments}. The same approach may well be
successful also in non-degenerate systems.

I dedicate this paper to the memory of my father, the physicist J. J. M.
Beenakker (1926--1998). Discussions with Oleg Bulashenko, Eugene Mishchenko,
Henning Schomerus, and Gilles Vissenberg are gratefully acknowledged. This work
was supported by the Dutch Science Foundation NWO/FOM.

\end{multicols}

\begin{references}
\bibitem{note3} A more accurate treatment including the diffusion term would
cut off the singularity at a value set by the charge density in the contact. As
a result, the electric field in the high-$V$ regime would not vanish at $x=0$
but extrapolate to zero at a point $x=-x_{\rm c}$ inside the contact.
Corrections to the shot-noise power of order $x_{\rm c}/L$ are estimated to be
less than 10\% in the computer simulations of Ref.\ \protect\cite{Gon98}.
\bibitem{note2} Thermal noise in a conductor with a non-linear $I$--$V$
characteristic is given by $P=4kT(V/\bar{I})(d\bar{I}/dV)^{2}$. Since
$\bar{I}\propto V^{2}$ in the space-charge limited regime, one finds
$P=8kTd\bar{I}/dV$ --- two times larger than the thermal noise in an Ohmic
conductor. This result can be derived from the theory presented in the text, by
invoking the thermal equilibrium condition
$\partial\bar{f}/\partial\varepsilon=-\bar{f}/kT$. Eq.\ (\protect\ref{Pbarf})
then takes the form $P=36\mu
kTL^{-2}\int_{0}^{L}dx\,(1-\sqrt{x/L})^{2}\bar{\rho}(x)$, which can be
evaluated using Eq.\  (\protect\ref{MottGurney}).
\end{references}
\end{document}